\documentstyle[11pt]{article}
\begin {document}
\begin{center}
{\bf ON THE STUDY OF THE SUBSTRUCTURE OF GALAXY CLUSTERS:
S-TREE TECHNIQUE IN NON-POINT APPROXIMATION
}
\vspace{0.1in}

K.M.Bekarian and A.A.Melkonian
\vspace{0.1in}

Department of Theoretical Physics, Yerevan Physics Institute,
Yerevan 375036, Armenia

\end{center}
\vspace{0.2 in}

\section{Introduction}

The study of substructuring properties of clusters of galaxies is one of the
basic problems on the way of understanding of the mechanisms of
formation of the large scale structure of the Universe. It is determined
by the fact that, the dynamical time scales of clusters of galaxies are
comparable with their ages, and hence their hierarchical properties
do contain essential primordial information.

A number of statistical methods are developed for the investigation of the
substructures of galaxy clusters \cite{Peeb, gk, Coles}.
Most of the methods are based on the analysis of the
positional information, combined with the use of redshift information. For
example,
wavelets (see e.g.\cite{Esc}) are usually used to analyze the data on
2D coordinates of galaxies, though the wavelets can be applied
for 1D redshift information as well. However such procedure is not
self-consistent and can lead to various biases.

From this point of view the S-tree technique developed by Gurzadyan,
Harutyunyan and Kocharyan
(see\cite{ghk, gk}) is using in a self-consistent
way the positional, kinematical and magnitude information about the system by
means of consideration of the phase space of the N-body systems and
revealing its typical properties.
S-tree method has been applied already for the study of substructure
of the Local Group\cite{GKP}, the core of Virgo cluster\cite{Petr}, and
Abell clusters of ESO Key Program survey (ENACS)\cite{GM}.

The existing version of S-tree\cite{gk} is based on the 'point'
approximation of the $N$-body system, i.e. it neglects the dimensions
of galaxies. However, as distinct to stellar
dynamics, when the point approximation is usually quite a fair
one, the `non-point' features of galaxies can be ocassionally
important in the study of clusters of galaxies.

Below we will inquire into the development of S-tree technique, in
non-point approximation, namely, considering the galaxies as
spheres of given radius, so that the cluster should appear as a set of
$N$ spheres.

Our second aim is the description of a refined algorithm of
analysis of the numerical output information of S-tree.

\section{S-tree technique}

First, let us briefly summarize the basics of S-tree technique;
for details we refer to the original papers\cite{gk, bg} and to
earlier references therein.

Consider $N$-body system as $X$ set.
$
  X= \{x_1,x_2,\dots,x_N\},
$
where $\forall x_i,\quad  (i=1,\dots,N)$ are given as follows

$L_i(l_i^1,l_i^2,l_i^3)$ \quad $L_i$ is coordinates vector;

$V_i({v_i}^{1},{v_i}^{2},{v_i}^{3})$ \quad $V_i$ is velocity vector

and $M_i$ are the masses of particles.

\vspace{0.1in}
Consider the following  $P$ function :

$ P: X\times X\rightarrow R_+ $ and $\rho\in R_+$.
\vspace{0.1in}
Now we give the basic definitions.

{\bf Definition 1.1}

We will say, that $\forall x \in X$ and $\forall y \in X$ are
$\rho$-bound, if:
$$
P(x,y)\geq\rho.
$$

{\bf Definition 1.2}

We will say, that $U \subset X (U \neq \emptyset)$ is $\rho$-bound
group, if:

$ a.\quad \forall x \in U$
and $\forall y \in X \setminus U \Rightarrow P(x,y)<\rho$

$b.\quad \forall x \in U$ and $\forall y \in U$
$\exists x=x_{i_1},x_{i_2},\dots,x_{i_k}=y,$

such
$P({x_i}_l,{x_i}_{l+1}) \geq \rho \quad  \forall l=1,\dots,k-1.$

{\bf Definition 1.3}

We will say, that $U_1,\dots,U_d$ is the distribution of set $X$ by
$\rho$-bound groups, if:

$a.\quad \bigcup_{i=1}^{d}U_i=X$

$b.\quad  i\neq j \quad (i=1,\dots, d; j=1,\dots, d) \Rightarrow U_i\bigcap
U_j=\emptyset$

$c.\quad  \forall U_d \quad (i=1,\dots,d)$ is
$\rho$-bounded group.
\vspace {0.1in}

Note that any $P$ function can be represented as $A
(a_{ij})$ matrix \cite{bg}.

Thus, we can have a representation of any $N$-body system via $S$-tree
diagram algorithm for given $A$ matrix and $\forall \rho$.
This distribution will satisfy the definitions 1.1,1.2 and 1.3.
This algorithm is efficient for any given $\rho$, so that one will
obtain the evolution of dependence of groups on $\rho$. We will represent the
final results in the form of tree-graph (S-tree) or a table.
The way one has to choose the final distribution is described below.

\section{Distribution via D matrix}

Consider again the set $X$. Via $\bar M_i$ let us denote the projection
of $M_i$ on $l^1l^2$ plane. Introduce the following set:
\vspace{0.1in}
$
Q=\{Q_1(\bar M_1,R_1),\dots,Q_N(\bar M_N,R_N)\},
$

where $R_i \in R_+$, $i=1,\dots,N$ and

$
Q_i(\bar M_i,R_i)=\{(l^1,l^2)|(l^1-l_i^1)^2+(l^2-l_i^2)^2 \leq
R_i^2\}.
$

Consider $\tilde P$ function, as:

$\tilde P: X  \rightarrow Q, \quad  \tilde P(x_i)=Q_i(\bar M_i,R_i)$

For $P_{\tilde p}: Q \times Q \rightarrow R_+$ and $\forall \rho
\in R_+$
we will give definitions, which are equivalent to definitions 1.1-1.3.

{\bf Definition 2.1}
We will say, that $\forall x_i \in  X$ and  $\forall x_j \in X$ are
$\rho$-bound,  if:
$$P_{\tilde p}(Q_i,Q_j)  \geq \rho, (i  \neq j)$$.

{\bf Definition 2.2}
We will say, that $U \subset X (U \neq \emptyset)$ is $\rho$-bound
group, if:

$ a. \quad \forall x_i \in U, \forall x_j \in \bar U \Rightarrow
P_{\tilde p}(Q_i,Q_j)<  \rho$

$b. \quad \forall x_i \in U,  \forall x_j \in U  (i \neq j)$
$\exists x_i=x_{i_0},x_{i_1},\dots,x_{i_k}=x_j, \forall t \in
(0,\dots  k-1)$

$$ P_{\tilde p}(Q_{i_t},Q_{i_t+1})  \geq \rho $$.

{\bf Definition 2.3}
We will say, that $U_1,\dots, U_d$ is the distribution of the
$\rho$-bound groups of the set $X$, if:

$a. \quad \bigcup_{i=1}^d U_i=X $

$b. \quad U_i  \bigcap U_j =  \emptyset$, if $i \neq j, \forall
i,j=1,\dots,d$

$c. \quad \forall U_i \quad i=1,\dots, d$ is $\rho$-bounded group.

Note, that $P_{\tilde p}(Q_i,Q_j)=0 \quad \forall i$.

Obviously, any $P_{\tilde p}$ function can be
represented as  $D(d_{ij})$ matrix. Now we will show a way of
construction of the matrix $D$.

If $i=j \quad \forall i,j=1,\dots,N \quad d_{ij}=0 $.

If $i  \neq j$

$$ d_{ij}=\left\{ \begin{array}{rcl}
                 0    & \mbox{if $\bar Q_i \bigcap \bar Q_j=
                         \emptyset$}\\
                 S_{Q_i \cap Q_j}& \mbox{otherwise}
                 \end{array}
        \right.
$$
where $\bar Q_i=Q_i \setminus \Gamma_i$, $\Gamma_i$-is bound  of
$Q_i$ and $S_{Q_i \cap Q_j}$ is the area of intersection of  $Q_i$
and $Q_j$. When we know $R_i$, $R_j$, $\bar M_i$, $\bar M_j$
values, it is easy to calculate the area $S_{Q_i \cap Q_j}$.

So, using $S$-tree diagrams algorithm, we can obtain the distribution
by $\rho$-bounded groups  of $X$ set according to defintions 2.1-2.3.

If we introduce also  the following function
$$
H=P_{\tilde p} \circ \hat P,
$$
where
$\hat P: X \times X \rightarrow Q \times  Q$.

$\hat P(x_i,x_j)=(\tilde{p}(x_i), \tilde
{p}(x_j))=(Q_i,Q_j),$
then we can use  only 1.1-1.3 definitions,
so long as
$$
H: X \times X \rightarrow R
$$.

\section{Output information algorithm}

The results of $S$-tree analysis can be  represented
in a form of a table \cite{ghk, gk}. For
optimization and simplification of the search of final distribution of
the system, i.e. the search of the necessary floor of the table,
one can use the following algorithm.

First, one has to perform a proper transformation of the latter table.
This will enable us to
obtain the information on the quantitative structure of each
distribution in a convenient form. This transition needs $M
\times  N$ actions, where $(M \times N)$ is $(a_{ij})$ matrix
dimension.
We correspond a distribution of $U_i$ numbers to each floor,
so that
$$U_i=\sum_{j=1}^Na_{ij}, \quad i=1,\dots,M.
$$
It is easy to see, that the first and the last floors of the
matrix $(a_{ij})$ are
$$
U_1=N
$$
$$
U_M=\sum_{j=1}^Nj;
$$
In some sense $U_i$  is reflecting the distribution 'density'
of the system's  subgroups. In order to decrease the
time  of the run of algorithm it is convenient to stop the investigation
of matrix at some $U_p$, where $p$ is the number of the floor with
$U_p  \geq  S$. One can choose, for example, as $x$ of
$U_M$, so that $x$ will depend on the matrix dimension.

By the next step we construct a chain of auxiliary matrices, in order
to obtain the final $K$ matrix reflecting the dynamics of
quantitative  and qualitative distribution of changes.

Each $i$th matrix is reflecting the quantitative dynamics of $i$th
subgroup. At the same time, $i$th matrix can be
represented as a function:
$$
f^*_i(\rho_t)=n^i_t,
$$
where $i=1,\dots,N$, $t=1,\dots,M$
and $n_i^k$ is the number of $i$th subgroup at $t$th floor.

   Note, that if we  'paste together' all mentioned auxiliary
matrices by 'floors', we will obtain exactly the sought-for $K$ matrix.

Now let us turn to the matrix $K$. We will consider  the
behavior of  the subgroup, which contains maximal  points $(m
\geq  2)$. We will fix  $\Delta \rho_i$ interval, i.e. $\Delta
\rho_i$-number of floor, when the group is preserving its  main
structure or if the group is loosing, for  example, the $x$
structure. Note, that $x$  has  to be defined for each matrix: if
the group is loosing more than $x$ of its numbers, then the process
will continue up to the next  floor.

Obviously, this process is finite, so far as the number
of subgroups $c_i$ is limited:
$$
c_1<c_i<c_M
$$,
where $c_1=1$ and $c_M=N$, i.e. the first floor contains all particles
from a single group (whole system), while on the last floor
each particle represents
a separate group. This readily follows from the Definitions 1.1-1.3  and
the method of choice of $\rho_t$.
Then, we have to choose the  maximal $\Delta  \rho_i$ and will take
the  corresponding distribution. As a final distribution we will
consider the floor, which results after the destruction of the mentioned
distribution.

For the realization of this algorithm we  need about $D=M^a \times
N^b$ actions \cite{kk}, where $a,b>0$.

Let us schematically illustrate this algorithm on a concrete example.
$
\left( \begin{array}{cccc}
A&A&A&A\\
A&B&A&A\\
A&B&A&C\\
A&B&C&D
\end{array}\right)
$
$\Rightarrow $
$
\left(\begin{array}{cccc}
1&1&1&1\\
1&2&1&1\\
1&2&1&3\\
1&2&3&4
\end{array}\right)
$

$
\left( \begin{array}{cccc}
1&2&3&4\\
1&&3&4\\
1&&3\\
1\\
\end{array}\right)
$      
$\circ $
$
\left(\begin{array}{cccc}
&\\
&2&&\\
&2&&\\
&2&&
\end{array}\right)
$
$\circ $
$
\left(\begin{array}{cccc}
&\\
&\\
&&&4\\
&&3
\end{array}\right)
$
$\circ $
$
\left(\begin{array}{cccc}
&\\
&\\
&\\
&&&4
\end{array}\right)
$

K=
$
\left(\begin{array}{cccccccccc}
1&2&3&4&*\\
1&&3&4&*&2\\
1&&3&&*&2&*&4\\
1&&&&*&2&*&3&*&4\\
\end{array}\right)
$

\section{Conclusion.}

We have proposed some new ways of development of S-tree
method in order to take into account the effects which could be important for
description of clusters of galaxies. Namely, we considered
the possibility of accounting for the non-zero
dimensions of the galaxies.
The described generalization of S-tree
can enable the construction of new type N-body models and can be
important especially while studying the denser core regions of galaxy
clusters.

We also represented an efficient procedure for the reliable
analysis of the output information of S-tree. As numerical
experiments, as well as the S-tree runs of the real galaxy clusters indicate,
this procedure can
become especially useful, when the numbers of galaxies in the studied
samples can become about 1000 and more. The obtaining of samples over
thousand of galaxies with known redshifts and magnitudes in the vicinity
of a given filament,
is within the aims of several forthcoming observational programs
(for reviews see Durret et al 1994).

We thank V.G.Gurzadyan and K.Oganessyan for valuable discussions.

\end{document}